# Superconductivity in correlated carbon nanotubes under pressure: A Bogoliubov-de Gennes study

Germán E. López and Chumin Wang*

*Instituto de Investigaciones en Materiales, Universidad Nacional Autónoma de México, CDMX, 04510, Mexico*

* Corresponding author. E-mail address: chumin@unam.mx

**Abstract**

In contrast to most microscopic theories of superconductivity based on the reciprocal space, the Bogoliubov-de Gennes (BdG) formalism provides a real-space alternative for addressing inhomogeneous systems. In this article, we study the superconducting states in correlated single-walled carbon nanotubes (SWNTs) with curvature and spin-orbit corrections, as well as the inter-tube interaction through a connecting molecule using an attractive Hubbard model. The results reveal a close relationship between the on-site superconducting gap and the single-electron local density of states. For the limiting case of independent large-diameter nanotubes, the BdG equations can be reduced to the standard Bardeen-Cooper-Schrieffer one with analytical solutions. Moreover, an optimal separation between nanotubes is found, which leads to a maximal superconducting critical temperature. This finding has a remarkable accordance with the experimental data obtained from Buckypapers built of boron doped SWNTs under external pressure.

## 1.- Introduction

Understanding the superconductivity with null electrical resistance and perfect diamagnetism has been a true challenge for the condensed matter physics since its discovery in 1911 [1]. The first microscopic theory of this strongly correlated many-body phenomenon was established in 1957 by J. Bardeen, L. Cooper and J. R. Schrieffer (BCS) proposing a multiple Cooper-pair wavefunction in reciprocal space originated from an attractive electron-electron interaction [2]. Shortly afterwards, N. Bogoliubov developed an alternative formalism based on the two-particle mean-field approximation and an analytical diagonalization procedure through unitary transformation. In 1964, P. G. de Gennes rewrote the Bogoliubov transformation in real space [3] and then, inhomogeneous superconductors with impurities and structural interfaces [4], as well as magnetic vortices in $d$-wave type-II superconductors [5] can be addressed.

Low-dimensional nanostructured materials constitute a dichotomy in the study of long-range coherent phenomena. On the one side, the reduction of kinetic energy as result of their low dimensionality could enhance the Cooper pairing for a given attractive electron-electron interaction. On the other side, the long-range coherent states, like superconductivity and ferromagnetism, in systems of one and two dimensions at finite temperatures are avoid by the Hohenberg-Mermin-Wagner theorem due to strong phase fluctuations in such systems [6]. Moreover, there is a size limit for the superconductivity in nanoparticles as stablished in the Anderson criteria, *i.e.*, the superconductivity could occur only when the electronic level spacing is smaller than the superconducting energy gap [7]. For the case of single-walled carbon nanotubes (SWNTs), proximity-induced supercurrents are measured when they are connected to superconducting electrodes [8], while the van Hove singularities of this almost one-dimensional material are exhibited by the zero-bias conductance peaks at low temperature [9]. In addition, their intrinsic superconductivity is observed in suspended ropes of SWNTs [10] or doped individual ones [11]. In fact, superconductivity at 15K has been reported in 4-Angstrom SWNTs embedded in aligned linear pores of the $AIOP_4$-5 zeolite [12].

The chirality and diameter of SWNTs determine their metallic or semiconducting behavior. Zigzag nanotubes have usually an energy gap around the Fermi level except for several specific radius presenting metallic properties, while armchair ones always are metallic [13]. The curvature of a small-diameter ($<1\,\text{nm}$) SWNT induces a new interaction between σ and π atomic orbitals producing a small energy gap in "metallic" zigzag SWNTs [14], as well as a relevant interaction between the electronic spin and their perimetral orbits on the nanotube [15]. These two interactions modify the hopping integrals [16,17] and may open electronic band gaps in both zigzag and armchair SWNTs [18], whose correction becomes crucial in small diameter nanotubes.

In this article, we study superconducting states of SWNTs by means of the Bogoliubov-de Gennes (BdG) equations based on an attractive Hubbard model and a unitary transformation presented in Appendix A, which reduces the BdG self-consistent equations of an armchair SWNT into standard BCS equations with analytical solution for the weak coupling limit. We further consider two parallel interacting SWNTs through a connecting molecule derived from the residual catalyst and the results reveal an optimal separation between nanotubes leading to a maximal superconducting critical temperature in accordance with experimental data obtained from Buckypapers built of boron doped SWNTs under pressure [19].

## 2.- The model

Let us consider an attractive Hubbard model given by [4,20-22]

$$\hat{H}_{\text{Hubbard}} = \sum_{j,s} \varepsilon_j \hat{n}_{j,s} + \sum_{<j,j'>,s,s'} t_{j,s;j',s'} \hat{c}^\dagger_{j,s} \hat{c}_{j',s'} + \sum_j U_j \hat{n}_{j,\uparrow} \hat{n}_{j,\downarrow}, \qquad (1)$$

where $\hat{c}^\dagger_{j,s}$ ($\hat{c}_{j,s}$) is the creation (annihilation) operator of electron with spin $s=\uparrow$ or $\downarrow$ at site $j$, $\hat{n}_{j,s} = \hat{c}^\dagger_{j,s}\hat{c}_{j,s}$, $\varepsilon_j$ is the on-site energy, $t_{j,s;j',s'}$ is the hopping integral between Wannier electronic states at nearest neighbors $j$ and $j'$ (denoted by $<j,j'>$) respectively with spins $s$ and $s'$, and $U_j$ is the on-site electron-electron interaction containing repulsive and attractive contributions correspondingly through exchanges of photon and phonon (or other bosons).

Applying the two-particle mean-field approximation (MF) to the electron-electron interaction term of Hamiltonian (1), it can be rewritten as

$$\hat{n}_{j,\uparrow}\hat{n}_{j,\downarrow} \approx \hat{c}_{j,\downarrow}\hat{c}_{j,\uparrow} <\hat{c}^\dagger_{j,\uparrow}\hat{c}^\dagger_{j,\downarrow}> + \hat{c}^\dagger_{j,\uparrow}\hat{c}^\dagger_{j,\downarrow} <\hat{c}_{j,\downarrow}\hat{c}_{j,\uparrow}> - <\hat{c}^\dagger_{j,\uparrow}\hat{c}^\dagger_{j,\downarrow}><\hat{c}_{j,\downarrow}\hat{c}_{j,\uparrow}>. \qquad (2)$$

The mean-field Hamiltonian for grand canonical ensembles ($\hat{H}_{\text{MF}} \approx \hat{H}_{\text{Hubbard}} - \mu\hat{N}$) with the chemical potential $\mu$ and the number operator $\hat{N} = \sum_{j,s} \hat{c}^\dagger_{j,s}\hat{c}_{j,s}$ takes the form

$$\hat{H}_{\text{MF}} \approx \hat{H} - \sum_j U_j <\hat{c}^\dagger_{j,\uparrow}\hat{c}^\dagger_{j,\downarrow}><\hat{c}_{j,\downarrow}\hat{c}_{j,\uparrow}>, \qquad (3)$$

where

$$\hat{H} = \sum_{j,s} (\varepsilon_j - \mu)\hat{n}_{j,s} + \sum_{<j,j'>,s,s'} t_{j,s;j',s'}\hat{c}^\dagger_{j,s}\hat{c}_{j',s'} + \sum_j (\Delta^*_j \hat{c}_{j,\downarrow}\hat{c}_{j,\uparrow} + \Delta_j \hat{c}^\dagger_{j,\uparrow}\hat{c}^\dagger_{j,\downarrow}) \qquad (4)$$

with $\Delta_j = U_j(<\hat{c}_{j,\downarrow}\hat{c}_{j,\uparrow}> - <\hat{c}_{j,\uparrow}\hat{c}_{j,\downarrow}>)/2$ obtained from the identity $\hat{c}_{j,\uparrow}\hat{c}_{j,\downarrow} = -\hat{c}_{j,\downarrow}\hat{c}_{j,\uparrow}$.

Using the Bogoliubov unitary transformation [23],

$$\begin{cases} \hat{c}_{j,\uparrow} = \sum_{\alpha \in \Xi} (u^\alpha_{j,\uparrow}\hat{\gamma}_\alpha - v^{\alpha*}_{j,\uparrow}\hat{\gamma}^\dagger_\alpha) \\ \hat{c}_{j,\downarrow} = \sum_{\alpha \in \Xi} (u^\alpha_{j,\downarrow}\hat{\gamma}_\alpha + v^{\alpha*}_{j,\downarrow}\hat{\gamma}^\dagger_\alpha) \end{cases}, \qquad (5)$$

with $\Xi = \{\alpha \mid E_\alpha \geq 0\}$, $[\hat{\gamma}_\alpha, \hat{\gamma}^\dagger_{\alpha'}]_+ = \delta_{\alpha,\alpha'}$ and $[\hat{\gamma}_\alpha, \hat{\gamma}_{\alpha'}]_+ = [\hat{\gamma}^\dagger_\alpha, \hat{\gamma}^\dagger_{\alpha'}]_+ = 0$, Hamiltonian (4) can be analytically diagonalized leading to [24]

$$\hat{H} = E_g + \sum_\alpha E_\alpha \gamma^\dagger_\alpha \gamma_\alpha, \qquad (6)$$

where $E_g$ is the ground state energy. Such diagonalization requires that amplitudes $u_{j,s}^\alpha$ and $v_{j,s}^\alpha$ of transformation (5) satisfy the Bogoliubov-de Gennes equation given by [23]

$$\Omega|\alpha\rangle = \begin{pmatrix} \mathbf{H}_{\uparrow,\uparrow} & \mathbf{H}_{\uparrow,\downarrow} & 0 & \Delta \\ \mathbf{H}_{\downarrow,\uparrow} & \mathbf{H}_{\downarrow,\downarrow} & \Delta & 0 \\ 0 & \Delta^* & -\mathbf{H}_{\uparrow,\uparrow}^* & \mathbf{H}_{\downarrow,\uparrow}^* \\ \Delta^* & 0 & \mathbf{H}_{\uparrow,\downarrow}^* & -\mathbf{H}_{\downarrow,\downarrow}^* \end{pmatrix} \begin{pmatrix} \mathbf{u}_\uparrow^\alpha \\ \mathbf{u}_\downarrow^\alpha \\ \mathbf{v}_\uparrow^\alpha \\ \mathbf{v}_\downarrow^\alpha \end{pmatrix} = E_\alpha \begin{pmatrix} \mathbf{u}_\uparrow^\alpha \\ \mathbf{u}_\downarrow^\alpha \\ \mathbf{v}_\uparrow^\alpha \\ \mathbf{v}_\downarrow^\alpha \end{pmatrix}, \quad (7)$$

where $(\Delta)_{j,j'} = \delta_{j,j'}\Delta_j$, $\mathbf{u}_s^\alpha = (u_{1,s}^\alpha, \cdots, u_{N,s}^\alpha)$, $\mathbf{v}_s^\alpha = (v_{1,s}^\alpha, \cdots, v_{N,s}^\alpha)$, elements $(j,j')$ of the Hamiltonian matrix $\mathbf{H}_{s,s'}$ are $(\mathbf{H}_{s,s'})_{j,j'} = t_{j,s;j',s'} + (\varepsilon_j - \mu)\delta_{s,s'}\delta_{j,j'}$, being $N$ the total number of atoms. In fact, the local superconducting gaps ($\Delta_j$) and local density of conducting electrons ($n_{j,s}$) per atom are related to the eigenvectors of equation (7) via

$$\Delta_j = -\frac{1}{2}U_j \sum_{\alpha \in \Xi} [u_{j,\uparrow}^\alpha v_{j,\downarrow}^{\alpha*} + u_{j,\downarrow}^\alpha v_{j,\uparrow}^{\alpha*}] \tanh\left[\frac{E_\alpha}{2k_B T}\right] \quad (8)$$

and

$$n_{j,s} = \langle \hat{c}_{j,s}^\dagger \hat{c}_{j,s} \rangle = \sum_{\alpha \in \Xi} \left\{ |u_{j,s}^\alpha|^2 f(E_\alpha) + |v_{j,s}^\alpha|^2 [1 - f(E_\alpha)] \right\}, \quad (9)$$

where $f(E_\alpha) = \{\exp[E_\alpha/(k_B T)] + 1\}^{-1}$ is the Fermi-Dirac distribution.

In a graphene sheet, carbon atoms are linked through σ bonds based on sp² hybrid orbitals occupied by three valence electrons of each atom, whose fourth one at $p_z$ orbital establishes a weak π bonds between nearest-neighbor atoms being the responsible of electronic transport in graphene, where the hopping integral between $p_z$ orbitals is $V_{pp}^\pi = -2.66$ eV [25,26]. In this article, we use a rectangular 4-atom unit cell, exhibited by green dashed lines in Figure 1(a), with translational vectors $\mathbf{a}_1 = (3a_{c\text{-}c}, 0)$ and $\mathbf{a}_2 = (0, a)$, where $a = \sqrt{3}a_{c\text{-}c}$ with $a_{c\text{-}c} = 1.42$ Å the bong length between nearest-neighbor carbon atoms. This graphene sheet may roll up to form an armchair SWNT of radius $R$ around the Y axis, as shown in Figure 1(b), where $\theta_j$ denotes the angle of $j$-th carbon atom in the cross section of a SWNT measured from the X axis.

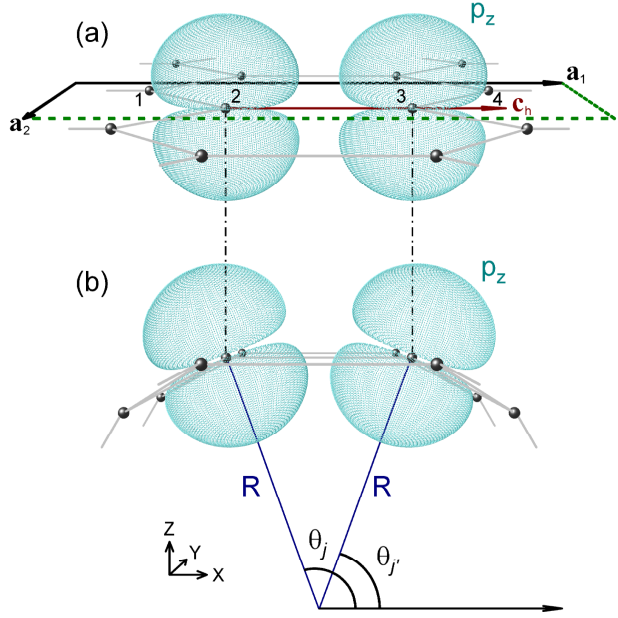

**Figure 1**. (Color online) Schematic representations of (a) a graphene sheet with the chiral vector $\mathbf{c}_h$, dashed green lines denoting the 4-atom unit cell with numbered atoms and translational vectors $\mathbf{a}_1$ and $\mathbf{a}_2$, while (b) a single walled carbon nanotube (SWNT) with the radius of nanotube $R$ and the angle of $j$-th atom $\theta_j$ with respect to the X axis.

In SWNTs, there are a rehybridization of $p_z$ and $sp^2$ orbitals due to the curvature (CV) as well as spin-orbital (SO) interaction derived from the electronic spin and its orbit motion along the cross section of SWNTs [15]. In consequence, the nearest-neighbor hopping integral $(t_{j,s;j',s'})$ in Hamiltonian (1) can be rewritten as [17]

$$t_{j,s;j',s'} = (t^{CV}_{j,s;j',s} + t^{SO}_{j,s;j',s})\delta_{s,s'} + t^{SF}_{j,s;j',\bar{s}}\delta_{s',\bar{s}}, \tag{10}$$

where

$$t^{CV}_{j,s;j',s} = V^\pi_{pp}\cos(\theta_j - \theta_{j'}) - (V^\sigma_{pp} - V^\pi_{pp})[1 - \cos(\theta_j - \theta_{j'})]^2 R^2/a^2_{c-c}, \tag{11}$$

$$t^{SO}_{j,\uparrow;j',\uparrow} = 2i\delta\sin(\theta_j - \theta_{j'})\left\{V^\pi_{pp} - (V^\sigma_{pp} - V^\pi_{pp})[\cos(\theta_j - \theta_{j'}) - 1]R^2/a^2_{c-c}\right\} = (t^{SO}_{j,\downarrow;j',\downarrow})^* \tag{12}$$

and

$$t^{SF}_{j,\uparrow;j',\downarrow} = (e^{-i\theta_j} + e^{-i\theta_{j'}})(V^\sigma_{pp} - V^\pi_{pp})[1 - \cos(\theta_j - \theta_{j'})]Y_{j',j}R\delta/a^2_{c-c} = -(t^{SF}_{j,\downarrow;j',\uparrow})^* \tag{13}$$

is the spin-flip (SF) term resulting also from the spin-orbital interaction. In equations (11), (12) and (13), $V^\sigma_{pp} = 6.38\, eV$ is the Slater-Koster σ-bond parameter [26], $\delta \approx 0.0028$ is the dimensionless spin-orbit interaction strength parameter [17] and $Y_{j',j} = Y_{j'} - Y_j$ is the distance between $j$ and $j'$ atoms along the Y axis.

For a periodic SWNT along the Y direction with a macroscopic length of $N_Y$ unit cells, subindex $j$ in equations (1)-(13) can be rewritten as $(q,l)$, where $q = 1, 2, \cdots, N_Y$ counts the cross-section cells (or perimetral belt) and $l = 1, 2, \cdots, N_a$ (with $N_a = 4N_0$) denotes the atoms in each cell, being $N_0$ the number of 4-atom unit cells in the cross-section cell. Thus, applying a Fourier transformation to the creation and annihilation operators, they become to

$$\begin{cases} \hat{c}_{j,s}^\dagger = \hat{c}_{q,l,s}^\dagger = (N_Y)^{-1/2} \sum_{k_Y}^{1BZ} \hat{c}_{l,s}^\dagger(k_Y) e^{-ik_Y(Y_q+Y_l)} \\ \hat{c}_{j,s} = \hat{c}_{q,l,s} = (N_Y)^{-1/2} \sum_{k_Y}^{1BZ} \hat{c}_{l,s}(k_Y) e^{ik_Y(Y_q+Y_l)} \end{cases}, \quad (14)$$

where $k_Y$ is a wave vector in the first Brillouin zone (1BZ), $Y_q$ and $Y_l$ are the Y-axis coordinates of $q$-th cross-section cell and of $l$-th atom in every cross-section cell, respectively.

Hence, Hamiltonian (4) becomes to

$$\hat{H} = \sum_{k_Y}^{1BZ} \{\hat{H}_0(k_Y) + \hat{H}_1(k_Y)\}, \quad (15)$$

where

$$\hat{H}_0(k_Y) = \sum_{l,s} (\varepsilon_l - \mu) \hat{c}_{l,s}^\dagger(k_Y) \hat{c}_{l,s}(k_Y) + \sum_{s,s',<l,l'>} t_{l,s;l',s'}(k_Y) \hat{c}_{l,s}^\dagger(k_Y) \hat{c}_{l',s'}(k_Y) \quad (16)$$

and

$$\hat{H}_1(k_Y) = \sum_l \left[ \Delta_l^* \hat{c}_{l,\downarrow}(k_Y) \hat{c}_{l,\uparrow}(k_Y) + \Delta_l \hat{c}_{l,\uparrow}^\dagger(k_Y) \hat{c}_{l,\downarrow}^\dagger(k_Y) \right]. \quad (17)$$

For example, the hopping integral term of $\hat{H}_0(k_Y)$ for a (1,1) SWNT containing only a single 4-atom unit cell in each cross section is given by

$$\sum_{<l,l'>} t_{l,s;l',s'}(k_Y) \hat{c}_{l,s}^\dagger(k_Y) \hat{c}_{l',s'}(k_Y) = \begin{cases} [2\cos\frac{k_Y a}{2}\delta_{s',s} + 2i\sin\frac{k_Y a}{2}\delta_{s',\bar{s}}] \\ \times [t_{1,s;2,s'} \hat{c}_{1,s}^\dagger(k_Y)\hat{c}_{2,s'}(k_Y) + t_{4,s;3,s'} \hat{c}_{4,s}^\dagger(k_Y)\hat{c}_{3,s'}(k_Y)] \\ + t_{1,s;4,s'} \hat{c}_{1,s}^\dagger(k_Y)\hat{c}_{4,s'}(k_Y) + t_{2,s;3,s'} \hat{c}_{2,s}^\dagger(k_Y)\hat{c}_{3,s'}(k_Y) + h.c. \end{cases}, \quad (18)$$

where $h.c.$ stands for Hermitian conjugated terms, $\bar{s}$ means the opposite spin with respect to $s$, and all the three corrections (10) are included in $t_{l,s;l',s'}$.

Moreover, the Bogoliubov transformation (5) can be rewritten as

$$\begin{cases} \hat{c}_{l,\uparrow}(k_Y) = \sum_{\alpha \in \Xi} [u_{l,\uparrow}^\alpha(k_Y)\hat{\gamma}_\alpha - v_{l,\uparrow}^{\alpha*}(k_Y)\hat{\gamma}_\alpha^\dagger] \\ \hat{c}_{l,\downarrow}(k_Y) = \sum_{\alpha \in \Xi} [u_{l,\downarrow}^\alpha(k_Y)\hat{\gamma}_\alpha + v_{l,\downarrow}^{\alpha*}(k_Y)\hat{\gamma}_\alpha^\dagger] \end{cases}, \quad (19)$$

where

$$\begin{cases} u_{l,\sigma}^{\alpha}(k_Y) = (N_Y)^{-1/2} \sum_{q=1}^{N_Y} u_{q,l,\sigma}^{\alpha} e^{ik_Y(Y_q+Y_l)} \\ v_{l,\sigma}^{\alpha}(k_Y) = (N_Y)^{-1/2} \sum_{q=1}^{N_Y} v_{q,l,\sigma}^{\alpha} e^{ik_Y(Y_q+Y_l)} \end{cases}. \quad (20)$$

Therefore, BdG equation (7) in the reciprocal space $(k_Y)$ can be written as

$$\Omega(k_Y)|\alpha(k_Y)\rangle = \begin{pmatrix} \mathbf{H}_{\uparrow,\uparrow}(k_Y) & \mathbf{H}_{\uparrow,\downarrow}(k_Y) & 0 & \Delta \\ \mathbf{H}_{\downarrow,\uparrow}(k_Y) & \mathbf{H}_{\downarrow,\downarrow}(k_Y) & \Delta & 0 \\ 0 & \Delta^* & -\mathbf{H}_{\uparrow,\uparrow}^*(k_Y) & \mathbf{H}_{\downarrow,\uparrow}^*(k_Y) \\ \Delta^* & 0 & \mathbf{H}_{\uparrow,\downarrow}^*(k_Y) & -\mathbf{H}_{\downarrow,\downarrow}^*(k_Y) \end{pmatrix} \begin{pmatrix} \mathbf{u}_{\uparrow}^{\alpha}(k_Y) \\ \mathbf{u}_{\downarrow}^{\alpha}(k_Y) \\ \mathbf{v}_{\uparrow}^{\alpha}(k_Y) \\ \mathbf{v}_{\downarrow}^{\alpha}(k_Y) \end{pmatrix} = E_{\alpha}(k_Y) \begin{pmatrix} \mathbf{u}_{\uparrow}^{\alpha}(k_Y) \\ \mathbf{u}_{\downarrow}^{\alpha}(k_Y) \\ \mathbf{v}_{\uparrow}^{\alpha}(k_Y) \\ \mathbf{v}_{\downarrow}^{\alpha}(k_Y) \end{pmatrix}, \quad (21)$$

where $\mathbf{H}_{s,s'}$ is a matrix of $N_a \times N_a$ elements given by $(\mathbf{H}_{s,s'})_{l,l'}(k_Y) = (\varepsilon_l - \mu)\delta_{s,s'}\delta_{l,l'} + t_{l,s;l',s'}(k_Y)$. From equations (8) and (9), the local superconducting gaps ($\Delta_l$) in equation (21) and the number of electrons in cross-section cell ($N_e$) are respectively

$$\Delta_l = -\frac{1}{2N_Y} U_l \sum_{k_Y}^{1BZ} \sum_{\alpha \in \Xi} [u_{l,\uparrow}^{\alpha}(k_Y) v_{l,\downarrow}^{\alpha*}(k_Y) + u_{l,\downarrow}^{\alpha}(k_Y) v_{l,\uparrow}^{\alpha*}(k_Y)] \tanh\left[\frac{E_{\alpha}(k_Y)}{2k_B T}\right] \quad (22)$$

and

$$N_e = \sum_{l=1}^{N_a} \sum_{s=\uparrow,\downarrow} n_{l,s} = \frac{1}{N_Y} \sum_{l=1}^{N_a} \sum_{s=\uparrow,\downarrow} \sum_{k_Y}^{1BZ} \sum_{\alpha \in \Xi} \{|u_{l,s}^{\alpha}(k_Y)|^2 f(E_{\alpha}) + |v_{l,s}^{\alpha*}(k_Y)|^2 [1-f(E_{\alpha})]\}, \quad (23)$$

where $N_a = 4N_0$ is the number of atoms in cross-section cell of an isolated SWNT. For the case of two correlated SWNTs with a connecting molecule located between these nanotubes, the number of atoms in cross-section cell is $N_a = 8N_0 + 1$.

In summary, starting from a uniform superconducting gap ($\Delta_l$), one obtains eigenvectors $\mathbf{u}_s^{\alpha}(k_Y)$ and $\mathbf{v}_s^{\alpha}(k_Y)$ from BdG equation (21) and a new set of $\Delta_l$ from (22), which will replace the initial uniform gaps and continue the calculation of eigenvectors and $\Delta_l$. This self-consistent calculation process finishes, when the difference between the initial and resulting $\Delta_l$ of a certain calculation step is less than the convergence criterion.

## 3.- Results

The single-electron dispersion relations $\varepsilon(k_Y)$ of an isolated SWNT can be obtained by solving the stationary Schrödinger equation with Hamiltonian $\hat{H}_0(k_Y)$ (16) using $\mu = 0$ and $\varepsilon_l = 0$. The results of $\varepsilon(k_Y)$ are plotted in Figures 2(a) for (2,2) and 2(b) for (10,10) armchair SWNTs with (red lines) and without (gray lines) curvature ($t_{j,s;j',s}^{CV}$), spin-orbital ($t_{j,s;j',s}^{SO}$) and spin-flip ($t_{j,s;j',\bar{s}}^{SF}$) corrections given in (11)-(13), using additionally $V_{pp}^{\pi} = -2.66$ eV, $V_{pp}^{\sigma} = 6.38$ eV, $\delta \approx 0.0028$ and $a_{c-c} = 1.42$ Å, while magnifications are exhibited in insets 2(a') and 2(a'') where green and blue lines are dispersion relations containing only the curvature correction and both curvature and spin-orbital contributions, respectively.

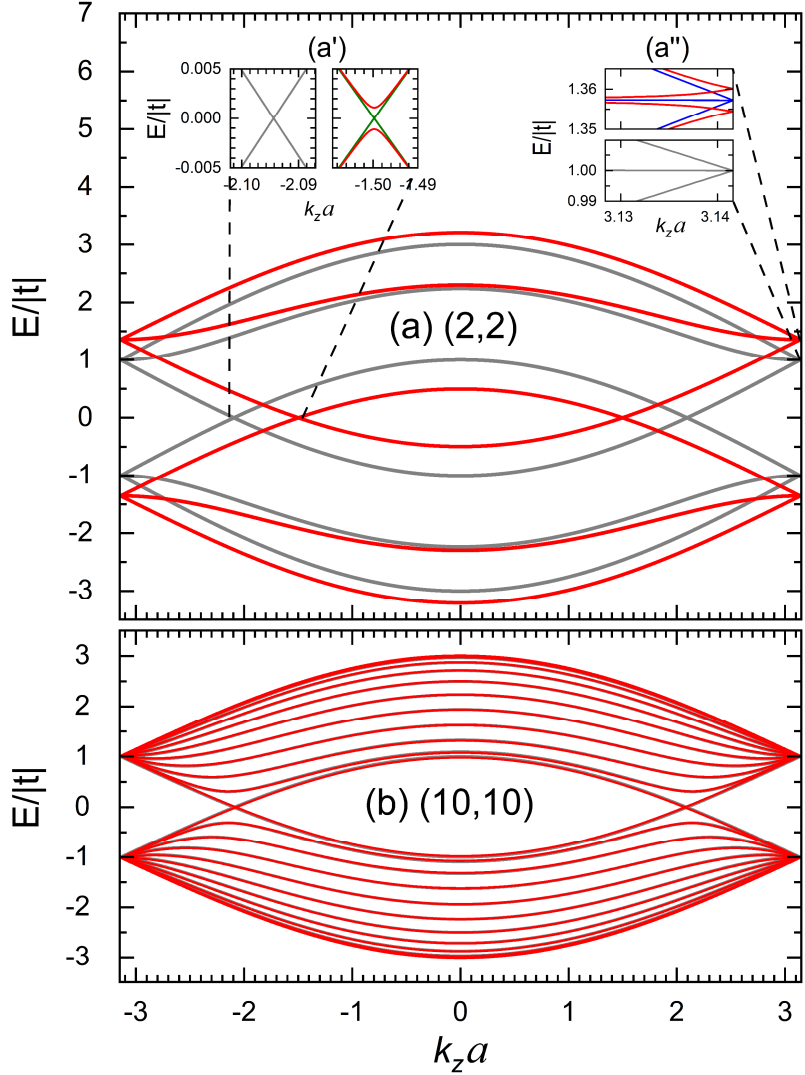

**Figure 2**. (Color online) Single-electron dispersion relations for (a) (2,2) and (b) (10,10) armchair SWNTs with (red lines) and without (gray lines) the curvature, spin-orbital and spin-flip corrections, while in insets (a') and (a'') the green lines include only the curvature correction and blue ones contain both curvature and spin-orbital contributions. The parameters used in this figure are given in the text.

Observe in Figure 2 that the curvature, spin-orbital and spin-flip corrections become remarkable when the radius of SWNTs reduces, and for a (10,10) SWNT with a diameter of 1.358 nm these corrections are very small, as mentioned in Ref. [27]. In fact, the central part around $E=0$ in Figure 2(b) reproduces Figure 1 of Ref. [18], where the spin-orbital correction was explicitly considered in its Hamiltonian.

The dispersion relations of Figures 2 without the spin-flip correction, $\varepsilon_n^\pm(k_Y)$ and $-\varepsilon_n^\pm(k_Y)$, can be analytically obtained as

$$\varepsilon_n^\pm(k_Y) = \sqrt{4|\tilde{t}_1|^2 \cos^2(k_Y) + |\tilde{t}_2|^2 \pm 4|(f_n + g_n)\cos(k_Y)|} \tag{24}$$

where

$$\tilde{t}_1 = V_{pp}^\pi - \left(\frac{7V_{pp}^\pi + V_{pp}^\sigma}{8}\right)\frac{a_{c-c}}{R}\left[\frac{a_{c-c}}{8R} - i\delta\sqrt{1 - \frac{a_{c-c}^2}{16R^2}}\right], \tag{25}$$

$$\tilde{t}_2 = V_{pp}^\pi - \left(\frac{V_{pp}^\pi + V_{pp}^\sigma}{2}\right)\frac{a_{c-c}}{R}\left[\frac{a_{c-c}}{2R} - 2i\delta\sqrt{1 - \frac{a_{c-c}^2}{4R^2}}\right], \tag{26}$$

$f_n = [\text{Im}(\tilde{t}_1)\text{Im}(\tilde{t}_2) - \text{Re}(\tilde{t}_1)\text{Re}(\tilde{t}_2)]\cos(n\pi/N_0)$ and $g_n = [\text{Im}(\tilde{t}_1)\text{Re}(\tilde{t}_2) + \text{Re}(\tilde{t}_1)\text{Im}(\tilde{t}_2)]\cos(n\pi/N_0)$, being $n = 1, 2, \cdots, N_0$ and $R = a_{c-c}\sqrt{5 + 4\cos[\pi/(2N_0)]}/\{4\sin[\pi/(2N_0)]\}$.

On the other hand, the normalized single-electron density of states (DOS) can be written as [28]

$$DOS(E) = -\frac{1}{N_a}\frac{a}{2\pi}\frac{1}{\pi}\lim_{\eta \to 0^+}\int_{-\pi/a}^{\pi/a} \text{Im}\{\text{Tr}[G(E + i\eta, k_Y)]\}\,dk_Y, \tag{27}$$

where $G(z, k_Y) = [z - \hat{H}_0(k_Y)]^{-1}$ is the retarded Green's function and the trace (Tr) is taken over the atoms of cross-section cell including the spin degrees of freedom. For an isolated SWNT with infinite length, there is a unique local superconducting gap ($\Delta$) for all atoms given by Eq. (22).

In Figures 3, (a, c) show the DOS containing an imaginary part of energy of $\eta = 0.0004\,|t|$ and (b, d) $\Delta$ with $U_l = -0.56|t|$ for isolated (a, b) (2,2) and (c, d) (8,8) armchair SWNTs. The results of both DOS and $\Delta$ without corrections are represented by gray lines and those with the curvature and spin-orbital corrections are plotted as blue lines, while the results with all the three curvature, spin-orbital and spin-flip improvements are drawn as red lines using the same Hamiltonian parameters of Figure 2. Notice the close correspondence between the DOS and $\Delta$ spectra, whose peaks are found at the same energies and a general broadening is observed in $\Delta$ spectra comparing with the DOS ones.

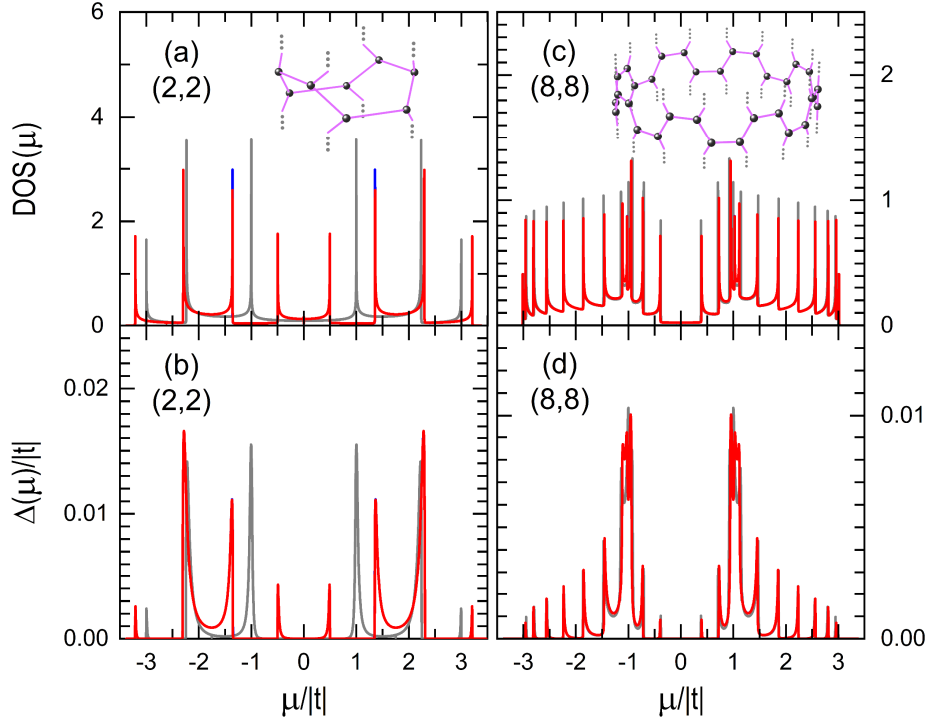

**Figure 3**. (Color online) (a,c) Normalized density of states (DOS) and (b,d) superconducting gap ($\Delta$) as functions of the chemical potential ($\mu$) for (a,b) (2,2) and (c,d) (8,8) SWNTs with $U_l = -0.56|t|$ and an imaginary part of energy $\eta = 4\times 10^{-4}|t|$. In (a-d), gray lines show results without corrections, blue lines include curvature and spin-orbital corrections, and red lines contain all the three curvature, spin-orbital and spin-flip corrections.

Note also the almost identical blue and red curves in Figures 3(c) and 3(d), *i.e.*, the spin-flip correction is truly small for both DOS and $\Delta$ spectra of (8,8) armchair SWNTs, which will be the system under study in the rest of this article. Hence, we assume $t^{SF}_{j,s;j',\bar{s}} = 0$, in consequence, there are simple expressions for the DOS and $\Delta$ given by [23]

$$DOS(E) = -\frac{1}{N_a}\frac{a}{2\pi}\frac{1}{\pi}\lim_{\eta\to 0^+}\int_{-\pi/a}^{\pi/a} dk_Y \sum_{n=1}^{N_0}\sum_{m=0}^{1} \text{Im}\left\{[E+i\eta-(-1)^m\varepsilon_n^+(k_Y)]^{-1} + [E+i\eta-(-1)^m\varepsilon_n^-(k_Y)]^{-1}\right\}, \quad (28)$$

and

$$1 = -\frac{a}{2\pi}\frac{U}{2N_a}\int_{-\pi/a}^{\pi/a} dk_Y \sum_{n=1}^{N_0}\sum_{m=0}^{1}\left\{\frac{\tanh[E^+_{n,m}(k_Y)/(2k_BT)]}{E^+_{n,m}(k_Y)} + \frac{\tanh[E^-_{n,m}(k_Y)/(2k_BT)]}{E^-_{n,m}(k_Y)}\right\}, \quad (29)$$

where $E^\pm_{n,m}(k_Y) = \sqrt{[(-1)^m\varepsilon_n^\pm(k_Y)-\mu]^2 + \Delta^2}$ and $\varepsilon_n^\pm(k_Y)$ are given by Eq. (24). In particular, the superconducting state of an isolated SWNT can be analyzed by means of the BCS formalism [2], as discussed in Appendix A, where equation (29) is deduced.

The superconducting critical temperature ($T_c$) can be calculated from equation (22) or (29) by taking $\Delta_l = \Delta = 0$ and $T = T_c$. In Figure 4, $T_c$ is plotted as a function of the attractive electron-electron interaction ($U_l = U$) and chemical potential ($\mu$) for isolated (8,8) SWNTs using the same parameter of Figures 3(c) and 3(d) without the spin-flip correction.

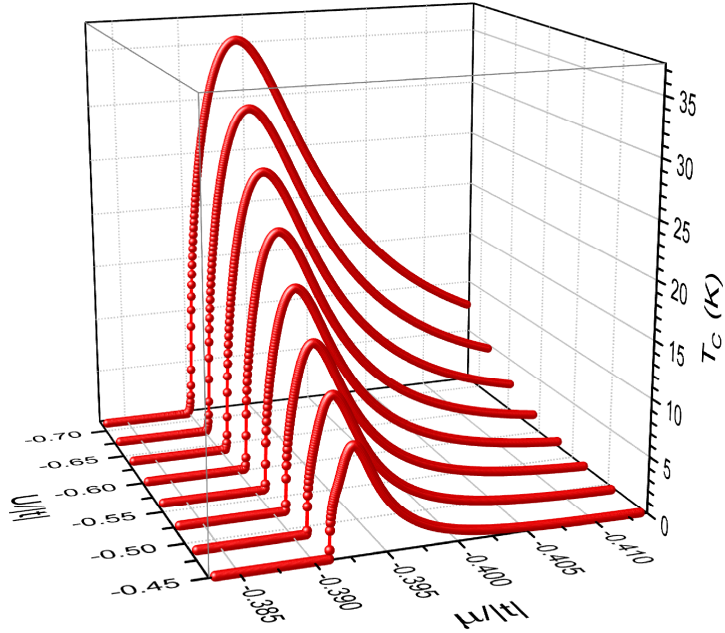

**Figure 4**. (Color online) Superconducting critical temperature ($T_c$) versus the electron-electron interaction ($U$) and chemical potential ($\mu$) for a (8,8) SWNT.

Given that isolated SWNTs are bipartite structures, the $\Delta$ spectra are symmetric with respect to $\mu = 0$, as shown in Figures 3. Hence, to address both slightly electron or hole dopped SWNTs, in Figure 4 we illustrate only a small section of $T_c$ spectrum around the nearest peak to $\mu = 0$ located around $\mu = -0.393 |t|$. Observe the growth of $T_c$ with $|U|$.

Now, let us consider a Buckypaper without rope formation [19], where the correlation between two SWNTs is considered through a set of independent residual-catalyst molecules characterized by a fictitious *s*-wave electronic state in each of them and all placed at the middle of two tubes, as shown in the sketch of a dual SWNT in Figure 5. The conducting electrons of SWNTs can hop between nanotubes passing through these molecule states with an on-site energy $\varepsilon_M$, an electron-electron correlation $U_M$ and hopping integrals $t'$ between the molecule and boundary carbon atoms.

The local superconducting gaps in this correlated dual SWNT are generally different and for the case of eight 4-atom unitary cells in the cross section of each SWNT, there are seventeen dissimilar superconducting gaps due to the mirror symmetries. These gaps are named $\Delta_l$ with $l = 0, 1, 2, \cdots, 16$, where zero denotes the molecule and other sixteen integers represent the half of atoms in each cross section of SWNT, as showed in the inset of Figure 5.

The local superconducting gaps ($\Delta_l$) as functions of the temperature ($T$) are presented in Figure 5 for the connecting molecule (red circle), the first (green pentagons), fourth (yellow square) and sixteenth (blue triangle) atoms of a dual (8,8) SWNT illustrated in the sketch of Figure 5, using $t'=-0.3|t|$ and $\mu = -|t|$, as well as $\varepsilon_C=0$ and $U_C=-|t|$ for carbon atoms, while $\varepsilon_M=-0.4|t|$ and $U_M=-0.1|t|$ for the connecting molecule. Notice that all the $\Delta_l$ tend to zero at the same critical temperature $k_B T_c = 0.0356|t|$, despite having different values at the zero temperature. Moreover, Figures 5(a)-5(d) show $\Delta_l$ versus the chemical potential ($\mu$) for $l = 0, 1, 4$ and $16$ in the same dual (8,8) SWNT. Observe the asymptotic behavior of atoms far away from the connecting molecule, whose $\Delta_l$ spectrum is like that of an isolated (8,8) SWNT.

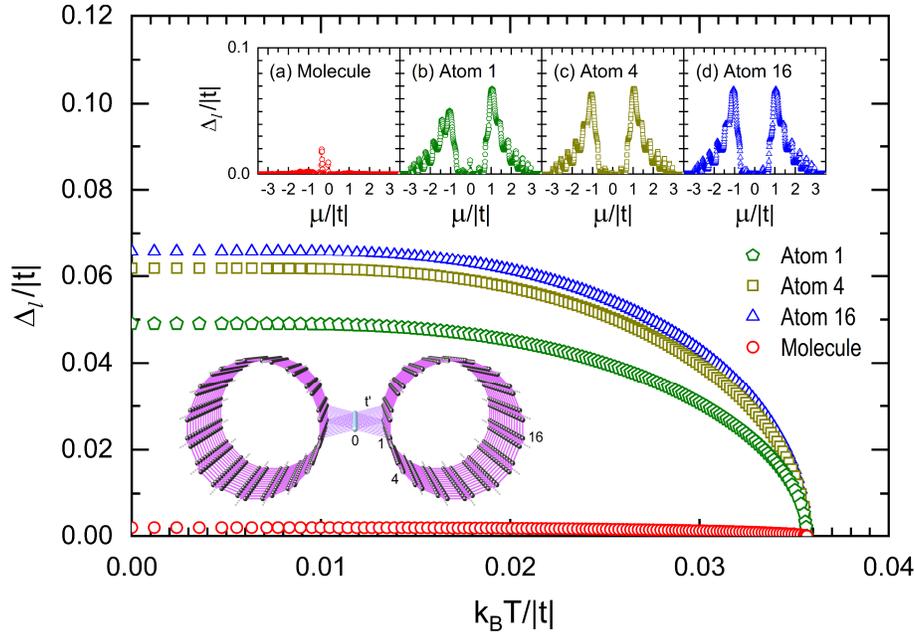

**Figure 5**. (Color online) Local superconducting gaps ($\Delta_l$) versus temperature ($T$) for sites $l=0$ (red circles), 1 (green pentagons), 4 (yellow squares) and 16 (blue triangles) of a dual (8,8) SWNT schematically illustrated in the inset, where the connecting molecule is numbered as site 0. Figures 5(a-d) show $\Delta_l$ as functions of $\mu$ for the mentioned sites.

Figure 6(a) presents the density of states as a function of the single electron energy ($E$) with an imaginary part $\eta = 10^{-3}|t|$ and the hopping integral ($t'$) between the molecule and SWNTs, for a correlated dual (8,8) SWNT with on-site energies $\varepsilon_C=0$ for carbon atoms and $\varepsilon_M=-0.4|t|$ for the connecting molecule. Observe in Figure 6(a) the main degenerate peak located at $\varepsilon_M$ derived from isolated molecules when $t'=0$ and the van Hove singularity at $E=-0.393|t|$ originated from isolated SWNTs, which are split into five peaks and one of them is more sensitive with $t'$ than others traveling from $E=-0.4|t|$ to $-0.02|t|$ when $t'$ changes from zero to $-0.3|t|$.

Given that the molecule-SWNT hopping integral ($t'$) is sensitive to the tube-molecule distance, it can be easily modified by the external pressure ($\sigma$). This relationship between $t'$ and $\sigma$ may be modeled as [29]

$$t'(\sigma) = t'(0)\exp(q\sigma/Y), \quad (30)$$

where $Y \approx 0.8$ GPa is the Young modulus of Buckypaper [30], $t'(0) = -0.48$ eV is the molecule-tube hopping integral without the external pressure [29] and the constant $q \approx 7.42$ [29,31].

Moreover, the substitutional boron doping concentration ($N_B$) can be calculated via

$$N_B = 1 - \frac{N_e - 2}{N_a - 1}, \quad (31)$$

where $N_e$ and $N_a$ are respectively the numbers of conducting electrons and atoms in the cross-section cell of a correlated dual SWNT with a connecting molecule. Actually, in Eq. (31), $N_e - 2$ and $N_a - 1$ are respectively the number of conducting electrons and carbon atoms in the cross-section cell of the two SWNTs without the connecting molecule, assuming fully occupied electronic states with both spins of the connecting molecule for $\mu > \varepsilon_M$.

In Figure 6(b), we plot the superconducting critical temperature (red spheres) obtained from the BdG equations (21) and (22) versus the external pressure ($\sigma$) and the boron concentration ($N_B$) determined from equation (31) for a dual correlated (8,8) SWNT with $\varepsilon_C=0$ and $U_C=-0.33|t|$ for carbon atoms, while $\varepsilon_M=-0.4|t|$ and $U_M=-0.03|t|$ for the connecting molecule. Notice that the main peak in Figure 6(b) corresponds to the most $t'$- sensitive branch in Figure 6(a), whose smaller

peak is also appeared in Figure 6(b), given the close relationship between the local density of states, local superconducting gap, and critical temperature.

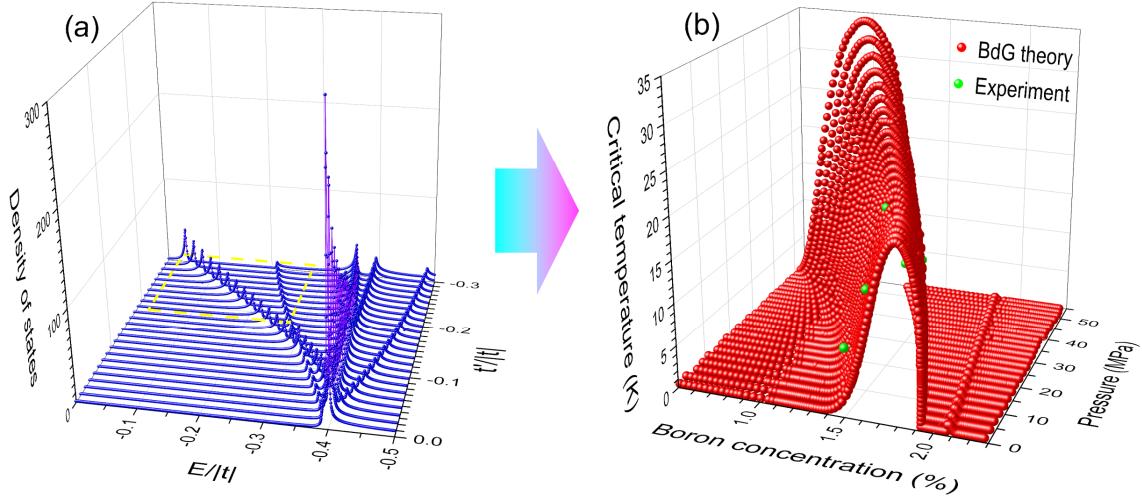

**Figure 6**. (Color online) (a) Density of states versus energy ($E$) and the hopping integral ($t'$) between the connecting molecule and SWNTs, as well as (b) superconducting critical temperature obtained from the BdG equations (red spheres) as a function of external pressure and boron concentration, both for a dual (8,8) SWNT shown in the inset of Figure 5. The section analyzed in Figure 6(b) is exhibited by yellow dashed lines in Figure 6(a), where the theoretical results are compared with experimental data (green spheres) obtained from a B-doped SWNT Buckypaper [19].

The theoretical results are compared in Figure 6(b) with the experimental ones obtained from a Buckypaper built of B-doped SWNTs with a diameter of 0.9 nm and a boron-doping concentration of 1.5% [19]. For the doping concentrations and external pressure analyzed in Figure 6(b), the corresponding zone (enclosed by yellow dashed lines) in Figure 6(a) is $E/|t| \in (-0.28, -0.03)$ and $t'/|t| \in (-0.3, -0.17)$, whose $\mu = E$ is larger than the on-site energy of connecting molecule $\varepsilon_M = -0.4|t|$ leading to two fully filled molecular electronic states with opposite spins when the connecting hopping integral $t' = 0$. In fact, the number of electrons $N_e$ in the dual SWNT is a function only of the boron concentration $N_B$, regardless the external pressure. Observe finally in Figure 6(b) the good agreement between the theoretical (red spheres) and experimental (green spheres) data.

## 4.- Conclusions

The Bogoliubov-de Gennes (BdG) formalism provides a real-space alternative to study the superconducting state in inhomogeneous systems, such as nanowires and correlated nanotubes, where the local superconducting gaps ($\Delta_l$) are sensitive to the proximity effect [32] and all the local gaps vanish at the same critical temperature, as shown in Figure 5. There is a strong correlation between $\Delta_l$ and the local density of states (see Figure 3), in accordance to the BCS analytical relationship $\Delta \approx 2\hbar\omega_c \exp\{-1/[N(0)V]\}$, where $\hbar\omega_c$ is the cutoff energy of attractive potential ($-V$) and $N(0)$ is the density of states (DOS) at the Fermi energy [33]. In addition, we have introduced the curvature and spin-orbit corrections into the BdG formalism causing an increment of density of states around 0.5|$t$| for small-diameter nanotubes. These corrections are less remarkable when the diameter is greater than 1 nm, as observed in Figure 3 in accordance with [27]. In Appendix A, we apply a unitary transformation to the BdG equations, which converts them to a BCS-type equation eliminating the self-consistent procedure. In other words, the unitary transformation reduces the $\Omega$ matrix of BdG equations into $N_a$ independent 2×2 matrices, whose analytical solutions are placed in the gap equation (22) converting it to the BCS equation.

However, the superconductivity of isolated SWNTs is essentially quenched by the long wavelength thermal fluctuations as prescribed by the Hohenberg-Mermin-Wagner theorem for one- and two-dimensional systems [34,35]. For a rope of SWNTs, the dimensionality restriction is overcome and the intertube interactions lead to a wider electronic band width due to the growth of average neighboring number, which increases the DOS at the Fermi energy located on the Dirac point or $E=0$ in Figure 3. In consequence, a superconducting critical temperature $T_C = 0.55$ K was reported for ropes of SWNTs [10]. Moreover, the rehybridization induced by the spin-orbit correction may have significant effects on the superconducting state of small-diameter carbon nanotubes, such as 4-Angstrom SWNTs in porous zeolite [36].

For the case of boron doped SWNTs (B-SWNT), the Fermi level may shift toward a van Hove singularity of the electronic DOS within the rigid band model causing a maximal $T_C$ of 12K in Buckypapers built of B-SWNTs [11]. This rigid band model has the advantage of being simple, providing a clear physical sense and good agreement with *ab-initio* results in the low doping limit [37,38]. The rope structure of B-SWNTs in Buckypapers causes a very small external-pressure effect on the superconducting state until 3 GPa [11], but these ropes can be dissolved by long-term centrifugate and ultrasonication. Hence, a new maximal $T_C$ of 19K is observed at an external pressure of 20 MPa [19]. Our results suggest that this external pressure causes the splitting of the van Hove singularity and connecting molecular states. One of these split bands eventually intersects with the Fermi energy producing a maximal $T_C$, as shown in Figure 6(b).

Finally, the BdG formalism including the curvature and spin-orbit interactions developed in this article could be useful for the study of superconducting states in small-diameter nanotubes, while the unitary transformation presented in Appendix A reduces the BdG equations of SWNT into the BCS ones without self-consistent procedure. The pressure induced splitting of the van Hove singularity and highly degenerate connecting-molecule state could be the origin of the maximum $T_C$ observed in Ref. [19]. The attractive Hubbard model used in this article can be improved via generalized Hubbard one without attractive density-density interactions [5], which enhances the superconducting ground state in competition with the charge-density wave (CDW), spin-density wave (SDW) and phase separation states [39]. On the other hand, the rigid band model utilized to calculate boron doping effects can also be improved by means of the virtual crystal approximation (VCA) or coherent potential approximation (CPA) [40,41], which is currently in progress.

**Acknowledgments**

This work has been partially supported by the Consejo Nacional de Humanidades, Ciencias y Tecnologías of Mexico (CONAHCyT) through grant CF-2023-I-830 and by the Universidad Nacional Autónoma de México (UNAM) through project PAPIIT-IN110823. Computations were performed at Miztli under the grant of LANCAD-UNAM-DGTIC-039. The technical assistances of Alejandro Pompa, Oscar Luna, Cain González, Silvia E. Frausto and Yolanda Flores are fully appreciated. G.E.L. acknowledges the doctoral fellowship from CONAHCyT.

**Appendix A. Unitary transformation from Bogoliubov-de Gennes to BCS equations**

The stationary Schrödinger equation for Hamiltonian $\hat{H}_0(k_Y)$ (16) without the spin-flip correction can be written as

$$\hat{H}_0(k_Y)\Psi_\upsilon(k_Y) = \xi_\upsilon(k_Y)\Psi_\upsilon(k_Y), \tag{A.1}$$

where $\upsilon = 1, 2, \cdots, N_a$ with $N_a$ the number of atoms in the cross-section cell of a single-walled carbon nanotube (SWNT) along the Y direction and

$$\begin{cases} \xi_{4n-3}(k_Y) = \varepsilon_n^+(k_Y) - \mu \\ \xi_{4n-2}(k_Y) = \varepsilon_n^-(k_Y) - \mu \\ \xi_{4n-1}(k_Y) = -\varepsilon_n^+(k_Y) - \mu \\ \xi_{4n}(k_Y) = -\varepsilon_n^-(k_Y) - \mu \end{cases}, \quad (A.2)$$

being $n = 1, 2, \cdots, N_0$ with $N_0$ the number of 4-atom unit cells in the cross section and then $N_a = 4N_0$. In Eq. (A.2), $\varepsilon_n^\pm(k_Y)$ are the dispersion relations given by Eq. (24).

The eigenfunctions of (A.1) can be written as $\Psi_\upsilon(k_Y) = \sum_\kappa w_{\kappa,\upsilon}(k_Y)|\kappa\rangle$, where $|\kappa\rangle$ is the Wannier function of $\kappa$-th atom in the cross section. There is a unitary transformation matrix ($\mathbf{W}$) built with eigenfunctions $\Psi_\upsilon(k_Y)$ written as column vectors, i.e.,

$$\mathbf{W}(k_Y) = \begin{pmatrix} w_{1,1}(k_Y) & \cdots & w_{1,N_a}(k_Y) \\ \vdots & \ddots & \vdots \\ w_{N_a,1}(k_Y) & \cdots & w_{N_a,N_a}(k_Y) \end{pmatrix}, \quad (A.3)$$

which diagonalizes the Hamiltonian $\hat{H}_0(k_Y)$ and their matrix elements satisfy the unitary conditions

$$\sum_{\upsilon=1}^{N_a} w_{\kappa,\upsilon}^*(k_Y) w_{\kappa',\upsilon}(k_Y) = \delta_{\kappa,\kappa'} \quad \text{and} \quad \sum_{\kappa=1}^{N_a} w_{\kappa,\upsilon}^*(k_Y) w_{\kappa,\upsilon'}(k_Y) = \delta_{\upsilon,\upsilon'}. \quad (A.4)$$

For an isolated SWNT, all the atoms are equivalent having the same superconducting gap ($\Delta$), i.e., the gap matrix ($\mathbf{\Delta}$) can be written as $\mathbf{\Delta} = \Delta \mathbf{I}$ with $\mathbf{I}$ the identity matrix. When the spin flip is neglected, the spin-up and spin-down components of Bogoliubov-de Gennes (BdG) equation (21) are separable and then, for each spin component

$$\mathbf{\Omega}(k_Y) = \begin{pmatrix} \mathbf{H}_0(k_Y) & \mathbf{\Delta} \\ \mathbf{\Delta}^* & -\mathbf{H}_0^*(k_Y) \end{pmatrix}, \quad (A.5)$$

can be transformed to

$$\tilde{\mathbf{\Omega}}(k_Y) = \begin{pmatrix} \mathbf{W}^\dagger(k_Y) & \mathbf{0} \\ \mathbf{0} & \mathbf{W}^\dagger(k_Y) \end{pmatrix} \mathbf{\Omega}(k_Y) \begin{pmatrix} \mathbf{W}(k_Y) & \mathbf{0} \\ \mathbf{0} & \mathbf{W}(k_Y) \end{pmatrix} = \begin{pmatrix} \tilde{\mathbf{H}}_0(k_Y) & \Delta \mathbf{I} \\ \Delta^* \mathbf{I} & -\tilde{\mathbf{H}}_0^*(k_Y) \end{pmatrix}, \quad (A.6)$$

where

$$\tilde{\mathbf{H}}_0(k_Y) = \mathbf{W}^\dagger(k_Y)\mathbf{H}_0(k_Y)\mathbf{W}(k_Y) = \begin{pmatrix} \xi_1(k_Y) & 0 & \cdots & 0 \\ 0 & \xi_2(k_Y) & 0 & 0 \\ \vdots & 0 & \ddots & \vdots \\ 0 & 0 & \cdots & \xi_{N_a}(k_Y) \end{pmatrix}, \qquad (A.7)$$

is a diagonal matrix. Given that the $\tilde{\mathbf{\Omega}}(k_Y)$ is formed by four diagonal submatrices [see Eq. (A.6)], $\xi_\upsilon(k_Y)$ is related only to $-\xi_\upsilon^*(k_Y)$ through superconducting gaps $\Delta$ and $\Delta^*$. Hence, $\tilde{\mathbf{\Omega}}(k_Y)$ can be rewritten as a 2×2 blocked matrix, whose BdG equation can be written as

$$\begin{pmatrix} \xi_\upsilon(k_Y) & \Delta \\ \Delta^* & -\xi_\upsilon(k_Y) \end{pmatrix}\begin{pmatrix} \tilde{u}_\upsilon(k_Y) \\ \tilde{v}_\upsilon(k_Y) \end{pmatrix} = E_\upsilon(k_Y)\begin{pmatrix} \tilde{u}_\upsilon(k_Y) \\ \tilde{v}_\upsilon(k_Y) \end{pmatrix}. \qquad (A.8)$$

The positive eigenvalues of (A.8) are

$$E_\upsilon(k_Y) = \sqrt{\xi_\upsilon^2(k_Y) + |\Delta|^2} \qquad (A.9)$$

and their eigenvectors are

$$\begin{cases} \tilde{u}_\upsilon(k_Y) = \dfrac{E_\upsilon(k_Y) + \xi_\upsilon(k_Y)}{\sqrt{|\Delta|^2 + [E_\upsilon(k_Y) + \xi_\upsilon(k_Y)]^2}} \\ \tilde{v}_\upsilon(k_Y) = \dfrac{\Delta^*}{\sqrt{|\Delta|^2 + [E_\upsilon(k_Y) + \xi_\upsilon(k_Y)]^2}} \end{cases}. \qquad (A.10)$$

Hence, equation (22) becomes to

$$\Delta = -\frac{a}{2\pi}\frac{U}{N_a}\int_{-\pi/a}^{\pi/a} dk_Y \sum_{\upsilon=1}^{N_a} \tilde{u}_\upsilon(k_Y)\tilde{v}_\upsilon^*(k_Y)\tanh\left[\frac{E_\upsilon(k_Y)}{2k_B T}\right]. \qquad (A.11)$$

Using (A.9) and (A.10), equation (A.11) can be rewritten as

$$1 = -\frac{a}{2\pi}\frac{U}{2N_a}\int_{-\pi/a}^{\pi/a} dk_Y \sum_{\upsilon=1}^{N_a} \frac{1}{E_\upsilon(k_Y)}\tanh\left[\frac{E_\upsilon(k_Y)}{2k_B T}\right], \qquad (A.12)$$

which is the well-known BCS equation. Equation (A.12) becomes to (29) by considering

$$E_\upsilon(k_Y) = E_{n,m}^{\pm}(k_Y) = \sqrt{[(-1)^m \varepsilon_n^{\pm}(k_Y) - \mu]^2 + \Delta^2}, \qquad (A.13)$$

where subindex $\upsilon = 1, 2, \cdots, N_a$ with $N_a = 4N_0$ includes three quantum numbers $n = 1, 2, \cdots, N_0$, $m = 0, 1$ and super index $\pm$.